%% file: icassp2024.tex
\documentclass{article}
\usepackage{spconf,amsmath,graphicx,xcolor,mathcomSTEv4}
\usepackage{mathtools}
\newcommand*{\Break}{\textbf{Break}}
\newcommand{\na}{\mathsf{n \backslash a}}

\input{preamble}

\title{Coding for the Unsourced B-Channel with Erasures: \\ Enhancing The Linked Loop Code}
\name{ William W. Zheng, Jamison R. Ebert, Stefano Rini, and Jean-Francois Chamberland
\thanks{
This material is based upon work supported, in part, by the National Science Foundation (NSF) under Grants CCF-2131106 and CNS-2148354, by Qualcomm Technologies, Inc., through their University Relations Program, and by the Ministry of Science and Technology (MOST) under Grant 110-2221-E-A49-052.
WWZ is with the Department of Information Engineering, the Chinese University of Hong Kong, HKSAR (email: wjzheng@link.cuhk.edu.hk).
JRE and JFC are with the Department of ECE
, Texas A\&M University, College Station, TX 77843, USA (emails: \{jrebert,chmbrlnd\}@tamu.edu).
SR is with the Department of ECE
, National Yang Ming Chiao Tung University (NYCU), Hsinchu, 300, TW (email: stefano.rini@nycu.edu.tw).
\textit{(Corresponding author: S. Rini.)}
}
}

\address{}

\begin{document}
%
\maketitle
\begin{abstract}
In \cite{zheng2023achan}, the linked loop code (LLC) is presented as a promising code for the unsourced A-channel with erasures (UACE). 
The UACE is an unsourced multiple access channel in which active users' transmitted symbols are erased with a given probability and the channel output is obtained as the union of the non-erased symbols.  
%
In this paper, we extend the UACE channel model to the unsourced B-channel with erasures (UBCE). 
The UBCE differs from the UACE in that the channel output is the multiset union -- or bag union-- of the non-erased input symbols. 
In other words, the UBCE preserves the symbol multiplicity of the channel output while the UACE does not.
%
Both the UACE and UBCE find applications in modeling aspects of unsourced random access.
%
%
The LLC from \cite{zheng2023achan} is enhanced and shown to outperform the tree code over the UBCE. 
Findings are supported by numerical simulations. 
\end{abstract}
\section{Introduction}
\label{sec:intro}
Massive machine-type communication (mMTC) is envisioned to play a significant role in next-generation wireless networks, where a colossal number of unattended, machine-type devices will sporadically seek to transmit short messages to one or more base stations. 
The rise of this class of network users poses a challenge to wireless infrastructure as traditional multi-user coordination processes are very inefficient under the envisioned traffic patterns.
%
As an alternative multiple access paradigm, unsourced random access (URA) was presented by Polyanskiy \cite{polyanskiy2017perspective} in which active users employ the exact same codebook and transmissions occur in a grant-free manner.

As motivation for this work, consider the problem of tracking shipping containers using satellite communications \cite{seung2005container}.
%
%
Due to the sheer number of containers and the need for energy efficiency, it is not feasible for the communication protocol to rely on a registration phase before resource allocation and transmission. 
Instead, it is more effective for the containers to send their data in an unsourced fashion. 
%
The satellite then receives the sum of all active containers'  updates at each transmission time. 
A higher protocol can then be employed to identify the route taken by each specific container, also accounting for the possibility of updates being lost due to channel imperfections and the ambiguity of connecting positional updates for a given container across time. 
%

%
%
%

For this and similar mMTC scenarios, concatenated coding techniques are often employed where an inner code protects against channel imperfections such as fading, asynchrony, and interference and an outer code is used for stitching transmissions together across time \cite{amalladinne2020coded}.
The effective channel as seen by the outer code may sometimes be modelled as the unsourced B-channel.
%
 

\noindent
{\bf Prior Work:}
The B-channel was introduced in \cite{chang1981t} as the $T$-user $M$-frequency channel with intensity information. 
In \cite{frolov2014capacity, bakin2016remark}, this channel is referred to as the S-user vector adder channel and its capacity under uncoordinated transmissions is investigated. 
Various coding schemes and their properties have been investigated in \cite{frolov2014coding, bakin2016sumrate}; however, we note that these schemes were not designed for an \textit{unsourced} B-channel. 
The unsourced paradigm was introduced in \cite{polyanskiy2017perspective} and the problem of coding for the unsourced A-channel (that is, the unsourced B-channel without intensity information) has been considered in~\cite{amalladinne2020coded, amalladinne2021unsourced, lancho2022finite, frolov2022ccslist}. 
In \cite{zheng2023achan}, the linked-loop code (LLC) is presented as a promising code for the unsourced A-channel with erasures (UACE). 
%

\noindent
{\bf Main contributions:}
In \cite{zheng2023achan}, the UACE is presented as an extension of the A-channel to the case when users' transmissions are encoded in an unsourced fashion and channel inputs may be erased with a fixed probability. 
A novel tail-biting graph-based coding scheme called the linked-loop code (LLC) is then presented as a promising code for the UACE. 
In this paper, we extend the UACE channel model to obtain the unsourced B-channel with erasures (UBCE), in which the channel outputs the multiset of non-erased input symbols. 
We enhance the LLC to obtain the eLLC and evaluate its performance over the UBCE. 
The eLLC outperforms the LLC in part due to its use of successive interference cancellation (SIC). 
As the UBCE models aspects of certain URA channels, the eLLC is seen as a promising code for improving the performance of URA schemes.

\section{Channel Model}
\label{sec:channel_model}

The UBCE is an extension of the B-channel to the case when all users employ the same codebook and each transmission is erased with a fixed i.i.d. probability. 
Let $\mathcal{Q}$, where $|\mathcal{Q}| = Q$, denote the input alphabet from which each user selects its coded symbols and let $\mathcal{Y}_{B}(\ell)$ denote the $\ell$th output of the B-channel.
Mathematically,
\begin{equation}
    \mathcal{Y}_{B}(\ell) = \biguplus_{k\in[K]} x_k(\ell),
\end{equation}
where $x_k(\ell) \in \mathcal{Q}$ is the $k$th user's (potentially coded) symbol at time $\ell$ and $\uplus$ denotes the multiset operation. 
Now, consider the erasure function
\begin{equation}
    \delta_{E}(x) = \begin{cases}
        x & \mathrm{if}~E=0 \\
        \emptyset & \mathrm{if}~E=1,\\
    \end{cases}
\end{equation}
where $E\stackrel{\text{iid}}{\sim}\mathcal{B}(p_e)$, $\mathcal{B}$ denotes the Bernoulli distribution, and $p_e$ is the fixed probability of symbol erasure. 
When all users employ an identical codebook, the $\ell$th output of the UBCE is given by 
\begin{equation}
    \mathcal{Y}(\ell) = \biguplus_{k\in[K]} \delta_{E_{k, \ell}}\left(x_k(\ell)\right).
\end{equation}
When transmissions are coded across $L$ channel uses, the corresponding $L$ channel outputs are given by 
\begin{equation}
    \mathcal{Y} = \left[\mathcal{Y}(0), \mathcal{Y}(1), \ldots, \mathcal{Y}(L-1)\right].
\end{equation}
Note that, unlike $|\mathcal{Y}_{B}(\ell)|$, $|\mathcal{Y}(\ell)|$ is a random variable due to the influence of the erasure operator. 
However, it is known that $0 \leq |\mathcal{Y}(\ell)| \leq K~~\forall \ell \in [L]$.

\subsection{Binary Vector Representation}

We are interested in the case when users' payloads are obtained from sequences of bits. 
Let $Q = 2^J$ for some $J \in \mathbb{Z}^+$ and let $\mathcal{Q} = \mathbb{F}_2^J$.
Thus, $x_k(\ell)$ may be expressed as the $J-$bit binary vector $\vv_k(\ell)$, where $\vv_k(\ell)$ consists of information section $\wv_k(\ell)$ and parity section $\pv_k(\ell)$. 
Throughout, we employ the notation $\xv.\yv$ to denote the (horizontal) concatenation of two vectors; thus, $\vv_k(\ell)=\wv_k(\ell).\pv_k(\ell)$. 
%
%
To emphasize the fact that we are dealing with sets of vectors, we denote the UBCE output by
\begin{equation}
    \Ybset(\ell) = \biguplus_{k \in [K]} \delta_{E_{k, \ell}}\left(\vv_k(\ell)\right).
\end{equation}
Hereafter, it will be convenient to represent user $k$'s $L$ binary vectors as a single row vector $\vv_k \in \mathbb{R}^{JL}$, where
\begin{equation*}
   \begin{split}
        \vv_k &= \vv_k(0).\vv_k(1)\ldots\vv_k(L-1) \label{v_k expression} \\
        &=\wv_k(0).\pv_k(0).\wv_k(1).\pv_k(1)\ldots\wv_k(L-1).\pv_k(L-1).
   \end{split}
\end{equation*}
The procedure for obtaining $\pv_k(\ell)$ depends on the outer code employed. 
The details of encoding and decoding of the linked loop code are provided in Section~\ref{sec:llc}.

\subsection{Performance Metrics for the UBCE}

Let $\Wbset \triangleq \{\vv_k : k \in [K]\}$ denote the set of codewords transmitted by the $K$ active users and let $\widehat{\Wbset}$ denote the receiver's estimate of $\Wbset$.
The receiver wishes to recover all transmitted codewords without erroneously recovering a codeword that was not transmitted.
These two objectives are measured using the following two metrics. 
\begin{definition}[Payload Dropping Probability]
    The \textit{payload dropping probability}, or PDP, is defined as 
    \begin{equation}
        P_{\textrm{PDP}} = \frac{1}{K}\sum_{k \in [K]}\mathbb{P}\left[\vv_k \notin \widehat{\Wbset}(\Ybset) | \vv_k \in \Wbset\right].
    \end{equation}
This metric quantifies the probability of the decoder not recovering a transmitted codeword.
\end{definition}
\begin{definition}[Payload Hallucination Probability]
    The \textit{payload hallucination probability}, or PHP, is defined as:
    \begin{equation}
        P_{\textrm{PHP}} = \f 1 {\Kh}  \sum_{k' \in [\Kh]} \Pr \lsb  \vv_{k'} \not \in \Wbset | \vv_{k'}  \in \widehat{\Wbset}(\Ybset)\rsb.
    \end{equation}
This represents the probability of the decoder recovering a codeword that was not actually transmitted by any active user. 
\end{definition}



\section{Enhanced Linked-Loop Code}
\label{sec:llc}

The LLC is presented in \cite{zheng2023achan} as a promising code for the UACE. 
In this section, we modify the LLC to operate over the UBCE by enhancing its decoder in various ways: 
(i) we allow for SIC to take place over multiple rounds 
(ii) we let decoding start from any section, and 
(iii) we provide flexibility in the coding rate. 
Given these changes from the original formulation in \cite{zheng2023achan}, we refer to this class of codes as the enhanced LLC (eLLC).

%
%

An $(LJ, RLJ)_{(M, J)}$ eLLC of rate $R$ is a tail-biting, binary, linear code that encodes $RLJ$ information bits into $L$ binary vectors of $J$ bits each. 
An eLLC codeword assumes the form
\begin{equation}
    \vv = \wv(0).\pv(0).\wv(1).\pv(1)\ldots\wv(L-1).\pv(L-1),
\end{equation}
where $\wv(\ell) \in \mathbb{F}_2^{b_l}$
denotes the information portion of section $\ell$, 
$\pv(\ell) \in \mathbb{F}_2^{p_l}$ 
denotes the parity portion of section $\ell$, and $b_l + p_l = J, \forall \ell \in [L]$. 
A defining characteristic of an eLLC is the fact that the parity bits in section $\ell$ depend only on the information bits in the previous $M$ sections, in a tail-biting manner. 
For each $\ell \in [L]$, consider $M$ generator matrices 
$\{\Gv_{\left([\ell-r-1]_L, \ell \right)} \in \mathbb{F}_2^{b_{[l-r-1]_L} \times p_l} : r \in [M]\}$, where $[x]_L$ denotes $x$ modulo $L$ and each matrix entry is the result of a Bernoulli trial with parameter $0.5$. 
Then, $\pv(\ell)$ is obtained as
\begin{equation}
\label{eq: coding rule}
    \pv(\ell) = \sum_{r\in [M]} \wv([\ell - r-1]_L)\Gv_{\left([\ell - r -1]_L, \ell \right)}.
\end{equation}
%
%
For example, if $L=16$ and $M=2$, the first three sets of parity bits are obtained as:
\eas{
\pv(0) & =  \wv(14)\Gv_{14,0} + \wv(15)\Gv_{15,0} \\
\pv(1) & =  \wv(15)\Gv_{15,1} + \wv(0)\Gv_{0,1} \\
\pv(2) & =  \wv(0)\Gv_{0,2} + \wv(1)\Gv_{1,2}. 
}
%
%
The decoding process is described in Algorithms~\ref{alg: eLLC decoding algorithm}~and~\ref{alg:stitch sections}.

\begin{algorithm}[t]
\caption{Pseudo code of the eLLC decoding algorithm. }
\label{alg: eLLC decoding algorithm}

\algrenewcommand\algorithmicrequire{\textbf{Input:}}
\algrenewcommand\algorithmicensure{\textbf{Output:}}

\begin{algorithmic}[1]
\Require UBCE output $\Yv=[\yv(0), \ldots , \yv(L-1)]$, $T$ maximum number of erasures to recover
\Ensure 
 A set of decoded messages $\widehat{\Wv}$

\Comment{Phase $0$: codewords suffering no erasures}

\State $\widehat{\Wv} \gets \emptyset$

 \State $\ell^{\dagger}=\arg\min_{\ell} |\yv(\ell)|$  

\State $\bold{Y}[0,1,...,L-1] \leftarrow \bold{Y}[\ell^{\dagger},\ell^{\dagger}+1,...,\ell^{\dagger}+L-1]$

\State {$\widehat{\Xv}_0 \gets \textbf{stitchSections}(\Yv,0)$}
\State {$\widehat{\Wv} \gets \textbf{extractInfoBits}(\Xhv_0)$}

\Comment{Phase $1$: stitching codewords with $j>0$  erasures}

\ForEach{$j \in [1:T]$ }

    \State $\Yv \gets \Yv \setminus \Xhv_{j-1}$
    
    \State $\ell^{\dagger}=  \arg\max_{\ell} |\yv(\ell)|$
    
    \State $\bold{Y}[0,1,...,L-1] \leftarrow \bold{Y}[\ell^{\dagger},\ell^{\dagger}+1,...,\ell^{\dagger}+L-1]$ 

    \State {$\widehat{\Xv}_j \gets \textbf{stitchSections}(\Yv,j)$}

    \State {$\widehat{\Wv} \gets \widehat{\Wv} \cup \textbf{extractInfoBits}(\widehat{\Xv}_j $})
\EndFor

\State \Return $\widehat{\Wv}$

\end{algorithmic}
\end{algorithm}

\begin{algorithm}[t]
\caption{Pseudo code of the $\textbf{stitchSections}(\Yv,j)$ subroutine}
\label{alg:stitch sections}

\algrenewcommand\algorithmicrequire{\textbf{Input:}}
\algrenewcommand\algorithmicensure{\textbf{Output:}}

\begin{algorithmic}[1]
\Require $\Yv$ channel output after $j-1$ rounds of SIC, $j$ minimum number of erasures per codeword.
\Ensure  $\Xhv_j$ estimated active codewords with $j$ erasures  
%
\State $\widehat{\Xv}_j \gets \emptyset$
    \ForEach{$x_0 \in \yv(0)$}
        \State $\Lsf_j(x_0) \gets \{x_0\}$
        \ForEach{$ \ell \in [1:L-1] $}
            \ForEach{$\uv \in \Lsf_j(x_0)$}
                \State $\Lsf_j(x_0) \gets \Lsf_j(x_0) - \uv $
            
                \ForEach{$x_\ell \in \yv(\ell)$}
                    \If{$\textbf{parityCheck}(\uv. x_\ell)$}
                        \State $\Lsf_j(x_0) \gets \Lsf_j(x_0) \cup \{ \uv. x_\ell \} $
                    \EndIf
                \EndFor

                \If{$\uv \text{ contains } <j $\text{ number of }$ \na $}
                    \State $\Lsf_j(x_0) \gets \Lsf_j(x_0) \cup \{\uv.\na \}$
                \EndIf
                
            \EndFor
        \EndFor
        \ForEach{$\uv \in \Lsf_j(x_0)$}
            \If{$\textbf{uniquelyDecodable}(\uv)$}
                \State $\vv \gets \textbf{uniquelyDecode}(\uv)$
                \State $\widehat{\Xv}_j \gets \widehat{\Xv}_j \cup \{\vv \}$; \Break
            \EndIf
        \EndFor
    \EndFor

\State \Return $\widehat{\Xv}_j$

\end{algorithmic}
\end{algorithm}
\noindent
With respect to Algorithm \ref{alg: eLLC decoding algorithm}:
\begin{itemize}[leftmargin=*,itemsep=-1.25 mm]
\item  {\bf [line 2]: } 
%
%
In phase $0$, we decode the messages that have no erased subsections. 
Decoding starts from the section with the smallest cardinality because that section is the most likely to contain zero-erasure codewords .
%
\item  {\bf [line 3]:} After the decoding root is chosen as $\ell^{\dagger}$,  the output is re-labelled to have the section $\ell^{\dagger}$ in position zero.
\item  {\bf [lines 4/5]:} the function $\textbf{stitchSections}$ is detailed in Algorithm \ref{alg:stitch sections}. 
The function $\textbf{extractInfoBits}$ extracts the information bits from the decoded codeword as in \eqref{v_k expression}.

\item  {\bf [lines 6]:} In phase $1$, codewords with erasures up to $T$ symbols are reconstructed.  

\item  {\bf [lines 7]:} The estimated codewords in the previous iterations are subtracted from $\Yv$. 

\item  {\bf [lines 8]:} 
%
In phase $1$, contrary to phase $0$, we choose the section with the highest cardinality.
This is due to computational considerations: if an erasure is present in the root during decoding, recovering this missing information is computationally expensive. 
%
%
\end{itemize}
With respect to Algorithm \ref{alg:stitch sections}:
\begin{itemize}[leftmargin=*,itemsep=-1.25 mm]
    \item  {\bf [line 2,4,8]:} For each element in the root section, we ``stitch'' elements in the following sections that match the parity check consistency through the function $\textbf{parityCheck}$
    \item {\bf [line 10-11,14]} 
    If 
    a sub-path contains $<j$ number of $\na$ symbols, then the decoder inserts an $\na$ symbol to be resolved. If the stitching process can finish, then the function $\textbf{uniquelyDecode}$ recovers the missing section values through the parity consistency with the other sections. 
\end{itemize}
%
%

Let us clarify further the function $\textbf{uniquelyDecode}$  with an example.
Consider a codeword $\vv$ which suffers only an erasure in the  $5$th section. In the decoding process, there exist a sub-path $\vv(0).\vv(1).\vv(2).\vv(3).\vv(4).\na .\vv(6).....\vv(L-1).$ which is parity consistent. 
To recover the missing information bits $\wv(5)$, we can use the \eqref{eq: coding rule} to setup the appropriate system of equations. For instance, for $M=2$ we have to solve
\eas{
\pv(6) & =  \wv(4)\Gv_{4,6} + \wv(5)\Gv_{5,6}, \\
\pv(7) & =  \wv(5)\Gv_{5,7} + \wv(6)\Gv_{6,7}.
}


\begin{remark}[ \bf Innovation with respect to \cite{zheng2023achan}]
The main innovations of Algorithm~\ref{alg: eLLC decoding algorithm} over \cite{zheng2023achan} are: 
(i) the use of multiple rounds of successive interference cancellation (SIC) in phase $0$ of Algorithm \ref{alg:stitch sections}, 
(ii) the selection of the root section, and 
(iii) the flexibility to accommodate different rates and different generator matrices.  
\end{remark}

\begin{remark}[$\bigcup$ vs $\biguplus$]
A natural question is the difference in the channel output when considering unions vs bag unions in the presence of erasures. 
%
%
With the presence of erasures, $\Pr \lsb \mathcal{Y}_A = \mathcal{Y}_B \rsb$ is equal to the probability that all the non-erased symbols are different. Put it mathematically,
\begin{equation}
    \begin{split}
        & \Pr \lsb \mathcal{Y}_A = \mathcal{Y}_B \rsb = \\
        &\quad \sum_{m=0}^{K} \f { {Q \choose m } m!}{Q^m}  { K \choose m } (1-p_e)^m p_e^{K-m}.  
    \end{split}
\end{equation}
\end{remark}

\section{Simulation Results}
\label{sec:results}


In this section, we seek to numerically characterize the performance of the eLLC over the UBCE. 
\footnote{
The code used to simulate the eLLC is available at \url{https://github.com/williamzheng0711/linked-loop-code}.
}
Throughout this section, we fix $J = 16$ and vary the number of sections $L$, the window size $M$, and the probability of erasure $p_e$.
We also fix $T=1$, which implies that the maximum number of erasures the eLLC/LLC can correct is $1$. 
Throughout this section, we simulate multiple versions of the eLLC over the same randomly generated set of messages, codebooks, and erasure patterns, when applicable. 

\begin{figure}[t]
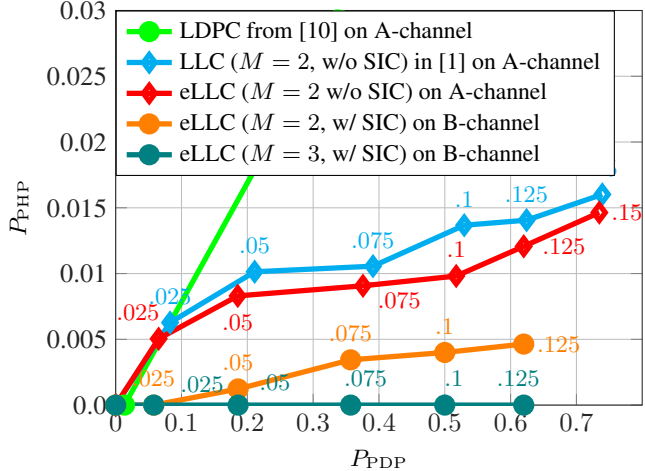

    \centering
    \include{Figures/canonical_changing_window_size}
    \vspace*{-6mm}
    \caption{Characterization of the performance of the eLLC in terms of $P_{\textrm{PHP}}$ and $P_{\textrm{PDP}}$ under $K=100, L=16, J=16$. Here, text labels on points denotes the probability of erasure $p_e$. }
    \label{fig:canonical_performance}
\end{figure}

Fig.~\ref{fig:canonical_performance} compares the performance of the eLLC/LLC with different window sizes on both the UACE and UBCE. 
In this figure, both the $x-$ and $y-$axes represent dependent variables which are functions of the independent variable $p_e$, which is provided by text labels over each point. 
As the goal is to minimize both $P_{\textrm{PHP}}$ and $P_{\textrm{PDP}}$, the best performing codes are those whose curves lie closest to the origin. 
The first comparison between the LLC and eLLC over the UACE clearly shows that the eLLC outperforms the LLC; this is due in part to the fact that any section can serve as a root section, thus providing additional decoding flexibility. 
We also note the reduced $P_{\textrm{PHP}}$ and $P_{\textrm{PDP}}$ achieved by the eLLC over the UBCE when compared to the UACE.
%
%
Finally, we note that the eLLC with $M=3$ curve lies below the eLLC with $M=2$ curve; this is because the additional parity structure offers a stronger discriminating power, which leads to a lower $P_{\textrm{PHP}}$.

\begin{figure}[t!]
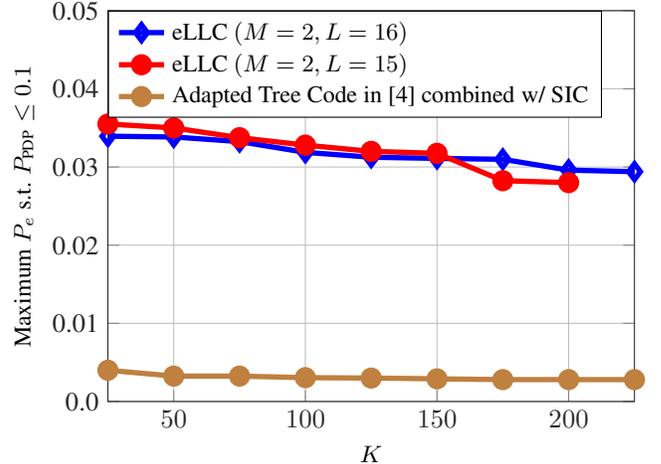

\centering
\include{Figures/maximal_affordable_pe}
\vspace*{-6mm}
\caption{Maximal $p_e$ s.t. $P_{\text{PDP}}\leq 0.1 $ vs. $K$. Tree code here is adapted to the B-channel setting combined with SIC.}
\label{fig: maximal_pe}
\end{figure}

In Fig.~\ref{fig: maximal_pe}, we study the maximum probability of erasure $p_e$ such that the eLLC can obtain a $P_{\textrm{PDP}} \leq 0.1$ as a function of the number of users $K$.
In this comparison, we vary $L$ to change the rate of the eLLC and also compare the eLLC with the contemporary tree code from \cite{amalladinne2020coded}. 
From Fig.~\ref{fig: maximal_pe}, it is clear that the eLLC outperforms the tree code!
We also note that the higher rate code ($L = 15$) outperforms the lower rate code ($L = 16$) for low number of users $K$ but that the low-rate code outperforms the high-rate code for a high number of users. 
%
%
%

\section{Conclusions}
\label{sec:conclusions}
In this paper, we discuss the problem of coding for the unsourced B-channel with erasures (UBCE).
The UBCE is an extension of the UACE in which the channel outputs the multiset, or set with multiplicities, of the input symbols. 
The linked loop code (LLC) from \cite{zheng2023achan} is adapted and enhanced for use over the UBCE. 
Numerical simulations show that the eLLC performs competitively over the UBCE.

\newpage
\bibliographystyle{IEEEbib}
\bibliography{IEEEabrv, icassp2024}

\end{document}

%% file: preamble.tex
\usepackage[utf8]{inputenc}

\usepackage{amsmath}
\usepackage{amsfonts}
\usepackage{amssymb}
\usepackage{textcomp}
\usepackage{color}
\usepackage{xcolor}
\usepackage{amsthm}
\usepackage{bm}
\usepackage{mathrsfs}
\usepackage{algorithmicx}
\usepackage{algorithm}
\usepackage[noend]{algpseudocode}
\usepackage{graphicx}
\usepackage{subcaption}
\usepackage{epsf}
\usepackage{xpatch}
\usepackage{setspace} 
\usepackage{enumitem,kantlipsum}
\usepackage{multirow,booktabs}
\usepackage{latexsym}
\usepackage{hyperref}
\usepackage{cleveref}
\usepackage[export]{adjustbox}
\usepackage{dsfont}
\usepackage{mathcomSTEv4}

\usepackage{tikz}
\usetikzlibrary{arrows,shapes,chains,matrix,positioning,scopes,patterns,calc,
decorations.markings,
decorations.pathmorphing,
}

\usepackage{pgfplots}
\pgfplotsset{compat=1.3}
\usepgflibrary{shapes}

\theoremstyle{definition}
\newtheorem{definition}{Definition}[section]

\newtheorem{remark}{Remark}

\newcommand{\Ybset}{\boldsymbol{\mathcal{Y}}}
\newcommand{\Wbset}{\boldsymbol{\mathcal{W}}}

\algnewcommand\algorithmicforeach{\textbf{for each}}
\algdef{S}[FOR]{ForEach}[1]{\algorithmicforeach\ #1\ \algorithmicdo}
\usetikzlibrary{decorations.pathreplacing,calc}

%% file: Figures/canonical_changing_window_size.tex
\begin{tikzpicture}

\begin{axis}[
    font=\small,
    width=7cm,
    height=5.25cm,
    scale only axis,
    every outer x axis line/.append style={white!15!black},
    every x tick label/.append style={font=\color{white!15!black}},
    xmin=0,
    xmax=0.8,
    xtick = {0.0, 0.1, ..., 0.8},
    xlabel={$P_{\mathrm{PDP}}$},
    xmajorgrids,
    every outer y axis line/.append style={white!15!black},
    every y tick label/.append style={font=\color{white!15!black}},
    ymin=0,
    ymax=0.03,
    ytick = {0.0, 0.005, 0.01, 0.015, 0.02, 0.025, 0.03},
    yticklabels={0.0, 0.005, 0.01, 0.015, 0.02, 0.025, 0.03},
    scaled y ticks=false,
    ylabel={$P_{\mathrm{PHP}}$},
    ymajorgrids,
    legend style={at={(0,1)},anchor=north west, draw=black,fill=white,legend cell align=left}
]

\addplot [
    color=green,
    solid,
    line width=2.0pt,
    mark size=3.0pt,
    mark=otimes*,
    mark options={solid}
]
table[row sep=crcr]{
    0.01333 0.0 \\
    0.337 0.02928 \\
    0.549 0.05252 \\
    0.692 0.077844 \\
    0.793 0.103896 \\
    0.876 0.15646  \\
};
\addlegendentry{LDPC from~\cite{amalladinne2021unsourced} on A-channel};

\addplot [
    color=cyan,
    solid,
    line width=2.0pt,
    mark size=3.0pt,
    mark=diamond,
    mark options={solid}
]
table[row sep=crcr]{ 
    0 0\\
    0.082631 0.006271\\
    0.211153 0.010135 \\
    0.390952 0.010561 \\
    0.529565 0.013673 \\ 
    0.624285 0.014058 \\
    0.739393 0.016018 \\
};
\addlegendentry{LLC ($M=2$, w/o SIC) in  \cite{zheng2023achan} on A-channel};

\node[label={90:{\color{cyan}$.025$}}] at (axis cs:0.082631, 0.006271) {};
\node[label={90:{\color{cyan}$.05$}}] at (axis cs:0.211153, 0.010135) {};
\node[label={90:{\color{cyan}$.075$}}] at (axis cs:0.390952, 0.010561) {};
\node[label={90:{\color{cyan}$.1$}}] at (axis cs:0.529565, 0.013673) {};
\node[label={90:{\color{cyan}$.125$}}] at (axis cs:0.624285, 0.014058) {};
\node[label={90:{\color{cyan}$.15$}}] at (axis cs:0.739393, 0.016018) {};

\addplot [
    color=red,
    solid,
    line width=2.0pt,
    mark size=3.0pt,
    mark=diamond,
    mark options={solid}
]
table[row sep=crcr]{
    0 0 \\
    0.06526 0.005042\\
    0.18545 0.008301 \\ 
    0.375715 0.009070 \\ 
    0.517391 0.009812 \\
    0.62  0.01207 \\
    0.734848 0.014639 \\
};
\addlegendentry{eLLC ($M=2$ w/o SIC) on A-channel};

\node[label={95:{\color{red}$.025$}}] at (axis cs:0.0826, 0.005042) {};
\node[label={-90:{\color{red}$.05$}}] at (axis cs:0.18545, 0.008301) {};
\node[label={300:{\color{red}$.075$}}] at (axis cs:0.375715, 0.01) {};
\node[label={90:{\color{red}$.1$}}] at (axis cs:0.517391, 0.009812) {};
\node[label={0:{\color{red}$.125$}}] at (axis cs:0.62, 0.01207) {};
\node[label={0:{\color{red}$.15$}}] at (axis cs:0.724848, 0.014639) {};

\addplot [
    color=orange,
    solid,
    line width=2.0pt,
    mark size=3.0pt,
    mark=otimes*,
    mark options={solid}
]
table[row sep=crcr]{
    0 0 \\
    0.05777 0 \\
    0.186 0.001227 \\ 
    0.357 0.00344 \\
    0.50 0.004 \\  
    0.62 0.00464 \\  
};
\addlegendentry{eLLC ($M=2$, w/ SIC) on B-channel};

\node[label={90:{\color{orange}$.025$}}] at (axis cs:0.05777, 0.0) {};
\node[label={90:{\color{orange}$.05$}}] at (axis cs:0.186, 0.001227) {};
\node[label={90:{\color{orange}$.075$}}] at (axis cs:0.357, 0.00344) {};
\node[label={90:{\color{orange}$.1$}}] at (axis cs:0.50, 0.004) {};
\node[label={0:{\color{orange}$.125$}}] at (axis cs:0.612, 0.00464) {};

\addplot [
    color=teal,
    solid,
    line width=2.0pt,
    mark size=3.0pt,
    mark=otimes*,
    mark options={solid}
]
table[row sep=crcr]{
    0 0 \\
    0.05777 0.0 \\
    0.186 0.0 \\ 
    0.357 0.0 \\ 
    0.50 0.0 \\ 
    0.62 0.0 \\ 
};
\addlegendentry{eLLC ($M=3$, w/ SIC) on B-channel};

\node[label={20:{\color{teal}$.025$}}] at (axis cs:0.07, 0.0) {};
\node[label={30:{\color{teal}$.05$}}] at (axis cs:0.19, 0.0) {};
\node[label={90:{\color{teal}$.075$}}] at (axis cs:0.38, 0.0) {};
\node[label={90:{\color{teal}$.1$}}] at (axis cs:0.5142, 0.0) {};
\node[label={90:{\color{teal}$.125$}}] at (axis cs:0.612, 0.0) {};

\end{axis}

\end{tikzpicture}

%% file: Figures/maximal_affordable_pe.tex
\pgfplotsset{scaled y ticks=false}
\begin{tikzpicture}

\begin{axis}[
    font=\small,
    width=7cm,
    height=5.2cm,
    scale only axis,
    every outer x axis line/.append style={white!15!black},
    every x tick label/.append style={font=\color{white!15!black}},
    xmin=25,
    xmax=225,
    xtick = {50, 100, ..., 225},
    xlabel={$K$},
    xmajorgrids,
    every outer y axis line/.append style={white!15!black},
    every y tick label/.append style={font=\color{white!15!black}},
    ymin=0,
    ymax=0.05,
    ytick={0.0, 0.01, 0.02, 0.03, 0.04, 0.05},
    yticklabels={0.0, 0.01, 0.02, 0.03, 0.04, 0.05},
    ylabel={Maximum $P_e$ s.t. $P_{\text{PDP}} \leq 0.1$},
    ymajorgrids,
    legend style={at={(0, 1)},anchor=north west, draw=black,fill=white,legend cell align=left}
]

\addplot [
    color=blue,
    solid,
    line width=2.0pt,
    mark size=3.0pt,
    mark=diamond,
    mark options={solid}
]
table[row sep=crcr]{ 
    25 0.03394 \\
    50 0.03384 \\
    75 0.03325 \\
    100 0.031875 \\
    125 0.03125 \\
    150 0.0311 \\
    175 0.030985 \\
    200 0.0296 \\
    225 0.0294 \\
};
\addlegendentry{eLLC $(M=2, L=16)$};

\addplot [
    color=red,
    solid,
    line width=2.0pt,
    mark size=3.0pt,
    mark=otimes*,
    mark options={solid}
]
table[row sep=crcr]{
    25 0.0355 \\
    50 0.035 \\
    75 0.03375 \\
    100 0.0328 \\
    125 0.032 \\
    150 0.03175 \\
    175 0.02825 \\
    200 0.028 \\
};
\addlegendentry{eLLC $(M=2,L=15)$};

\addplot [
    color=brown,
    solid,
    line width=2.0pt,
    mark size=3.0pt,
    mark=otimes*,
    mark options={solid}
]
table[row sep=crcr]{
    25 0.004 \\
    50 0.00325 \\
    75 0.00325 \\
    100 0.00305 \\
    125 0.003 \\
    150 0.0029 \\
    175 0.0028 \\
    200 0.0028 \\
    225 0.0028 \\
};
\addlegendentry{Adapted Tree Code in \cite{amalladinne2020coded} combined w/ SIC};

\end{axis}

\end{tikzpicture}